\documentstyle[preprint,aps]{revtex} 
\begin{document} 
\input epsf
\tighten

\title{Finite-size scaling for near-critical continuum fluids at \\ constant
pressure}

\author{N. B. Wilding and K. Binder}
\address{Institut f\"{u}r Physik, Johannes Gutenberg Universit\"{a}t, \\
Staudinger Weg 7, D-55099 Mainz, Germany.}

\setcounter{page}{0}
\maketitle 

\begin{abstract}

We consider the application of finite-size scaling methods to
isothermal-isobaric (constant-NpT) simulations of pure continuum fluids.  A
finite-size scaling {\em ansatz} is made for the dependence of the relevant
scaling operators on the particle number.  To test the proposed scaling
form, constant pressure simulations of the Lennard-Jones fluid at its
liquid-vapour critical point are performed.  The critical scaling operator
distributions are obtained and their scaling with particle number found to
be consistent with the proposed behaviour.  The forms of these scaling
distributions are shown to be identical to their Ising model counterparts. 
The relative merits of employing the constant-NpT and grand canonical
(constant-$\mu$VT) ensembles for simulations of fluid criticality are also
discussed. 

\end{abstract}
\thispagestyle{empty}
\newpage

\section{Introduction} 
\label{sec:intro}

The study of the phase behaviour of simple and complex fluids by
computer simulation is a subject of much current research activity
\cite{ALLEN2}. Of particular interest are the critical point properties of
such systems \cite{WILDING5}. In the vicinity of a critical point, the
correlation length grows extremely large and may exceed the linear size
of the simulated system.  When this occurs, the singularities and
discontinuities that characterise critical phenomena in the thermodynamic
limit are shifted and smeared out \cite{BINDER1}.  Unless care is exercised,
such finite-size effects can lead to serious errors in computer simulation
estimates of critical point parameters.

To cope with these problems, finite-size scaling (FSS) techniques have been
developed \cite{PRIVMAN,BINDER1}.  FSS methods enable one to extract
accurate estimates of infinite-volume thermodynamic quantities from
simulations of finite-sized systems.  To date, their application to fluid
criticality has been principally in conjunction with simulations in the
constant-$\mu$VT or grand-canonical ensemble (GCE).  The principal merit of
this ensemble is that the particle density fluctuates on the scale of the
system as a whole, thus enabling direct measurement of the large-scale
density fluctuations that are the essential feature of fluid criticality. 
The GCE has proven its worth in FSS studies of criticality in a variety of
fluid systems including the Lennard-Jones (LJ) fluid \cite{BRUCE2,WILDING}
and a 2D spin fluid model \cite{WILDING1}. 

Notwithstanding its wide utility, however, there exist many complex fluids
for which use of the GCE ensemble is by no means efficient.  Systems such as
semi-dilute polymer solutions are difficult to simulate in the GCE due to
excluded volume effects which hinder chain insertions.  While smart
insertion techniques go some way to ameliorating this difficulty, the long
chain lengths of greatest interest are currently inaccessible
\cite{WILDING4}.  Similarly, electrolyte models such as the restricted
primitive model show very poor acceptance rates for particle insertions due
to the formation of bound ion clusters \cite{MAKIS}.  Thus it is interesting
to ask whether one can deal with the near-critical density fluctuations in
such systems {\em without} having to implement inefficient particle transfer
operations. 

The approach we consider here, is to employ an ensemble wherein the total
particle number is fixed, but the density is allowed to fluctuate by virtue
of {\em volume} transitions.  Specifically, we consider how the FSS ideas,
hitherto only applied to systems with constant volume, may be generalised to
an isothermal-isobaric (NpT-ensemble) simulation.  Since finite-size scaling
usually rests on the idea of comparing the correlation length with the
(fixed) linear dimensions of the systems, the generalisation to systems
whose linear dimensions are dynamically fluctuating is not completely
obvious.  We make a scaling {\em ansatz} for the near-critical scaling
operator distributions and scaling fields, expressed in terms of powers of
the particle number.  This is then tested via a simulation study of the
critical Lennard-Jones fluid, in which it found that the FSS predictions are
indeed consistent with the simulation results.  Finally we discuss the
relative merits of the NpT- and $\mu$VT- (GCE) ensembles
for simulation studies of fluid criticality. 

\section{Finite-size scaling theory}

We consider a classical single component fluid, whose configurational energy
(which we write in units of $k_BT$) resides in a sum of pairwise interactions
amongst the $N$ particles it contains

\begin{equation}
\Phi (\{\vec{r} \}) = \sum_{i<j=1}^N\phi(\mid{\bf r}_i-{\bf r}_j\mid)
\label{eq:en}
\end{equation}
where $\phi(\mid{\bf r}_i-{\bf r}_j\mid)$ is the interparticle
potential which, for this work, we assign the Lennard-Jones form:

\begin{equation}
\phi(r)=4w[(\sigma/r)^{12}-(\sigma/r)^6]\label{eq:LJdef}
\end{equation}
where $w$ is the dimensionless well depth and $\sigma$ serves to set the
length scale.

Within the isothermal-isobaric ensemble, the partition function is given by
\begin{equation}
Z_N= \prod_{i=1}^N \left\{\int d\vec{r}_i\right\} \int dV
e^{ \left[ -\Phi (\{ \vec{r} \} ) -pV \right] }
\label{eq:Z}
\end{equation}
where $p=P/k_BT$ is the pressure, and $V=L^d$ is the homogeneously fluctuating system volume. 
The associated (Helmholtz) free energy is

\begin{equation}
G=-k_BT\ln Z_N= k_BT( \Phi (\{ \vec{r} \} )+pV)  
\label{eq:G}
\end{equation}

In the vicinity of the liquid-vapour critical point, the coarse-grained
properties of the system are expected to exhibit scaling behaviour.  For
simple fluids with short-ranged interactions, this behaviour is Ising-like
\cite{BRUCE2} and is controlled by two relevant scaling fields $\tau$ and
$h$ that measure deviations from criticality.  In general (in the absence of
the special Ising `particle-hole' symmetry), these scaling fields comprise
linear combinations \cite{REHR} of the reduced chemical potential $\mu$ and
well depth $w$:

\begin{equation}
\tau=w_c-w+s(\mu-\mu_c), \hspace{1cm} h=\mu-\mu_c+r(w_c-w)
\end{equation}
where subscripts denote critical values, and $s$ and $r$ are
non-universal field mixing parameters, the values of which depend on the
specific system under consideration. 

The respective conjugate operators are defined by \cite{BRUCE2}
\begin{equation}
\langle {\cal E} \rangle = \frac{1}{\langle V\rangle}\frac{\partial \ln Z_N}{\partial
\tau}, \hspace{1cm} 
\langle {\cal M} \rangle = \frac{1}{\langle V\rangle}\frac{\partial \ln Z_N}{\partial h}
\end{equation}
whereupon, utilizing equation~\ref{eq:Z},  and noting that  $dP=gdT+\rho d\mu$ (with
$g\equiv S/V$, the entropy density), one finds
\begin{equation}
{\cal E} =\frac{1}{1-sr}[u-r\rho], \hspace{1cm} 
{\cal M} =\frac{1}{1-sr}[\rho-su]
\end{equation}
where $\rho \equiv N/V$ is the particle density and $u\equiv \Phi(\{{\bf
r}\})/V$ is the energy density. We term ${\cal M}$ the ordering operator,
while ${\cal E}$ is termed the energy-like operator.

For the finite-size scaling behaviour of the joint distribution $P_N({\cal
M,E})$ we make the following {\em ansatz}:

\begin{equation}
P_N({\cal M,E})\simeq a^{-1}_{\cal M}a^{-1}_{\cal E}N^xN^y\tilde{p}(a^{-1}_{\cal
M}N^x({\cal M-M}_c),a^{-1}_{\cal E}N^y({\cal E-E}_c),
a_{\cal M}N^{1-x}h,a_{\cal E}N^{1-y}\tau )\\
\label{eq:ansatz}
\end{equation}
where $a_{\cal M}$ and $a_{\cal E}$ are non-universal metric factors and
$\tilde{P}$ is a universal function in the sense that it is identical for all
members of a given universality class {\em and} a given choice of boundary
conditions. Here we have chosen simply the particle number $N$ rather than
the volume $V$ as a measure of the `finite-size' of the systems, using now
suitable powers of $N$ in the scaling assumption.

The unspecified exponents $x$ and $y$ in equation~\ref{eq:ansatz} can be
related to the standard critical indices $\beta$ and $\nu$ by an argument
analogous to that invoked by Binder \cite{BINDER2} for the Ising model in
the canonical ensemble.  Consider the generalised isothermal
compressibility, $\kappa_\tau$, defined by

\begin{equation}
\kappa_\tau\equiv -\frac{1}{V}\left (\frac{\partial^2G}{\partial h^2}
\right )_\tau
\end{equation}
This may be reexpressed as a fluctuation relation by appeal to
equations~\ref{eq:Z} and ~\ref{eq:G}, whence one finds

\begin{equation}
\kappa_\tau=\beta \langle \rho\rangle^{-1}N \langle ({\cal M-M}_c)^2\rangle
\label{eq:kappa}
\end{equation}
From equation~\ref{eq:ansatz}, the scaling of the fluctuation in the
ordering operator ${\cal M}$ at finite $\tau$ is given by

\begin{eqnarray}
\langle({\cal M-M}_c)^2\rangle & = &a^{-1}_{\cal M}N^x\int d{\cal M}({\cal
M-M}_c)^2\tilde{p}_N(a^{-1}_{\cal M}N^x({\cal M-M}_c),N^{1-y}\tau)\\
&=& N^{-2x}a^2_{\cal M}\int dz z^2\tilde{p}_N(z,N^{1-y}\tau)\\
&\equiv& N^{-2x}a^2_{\cal M}\; f(N^{1-y}\tau)
\label{eq:scaleM}
\end{eqnarray}
where $f(N^y\tau)$ is an unspecified scaling function. 

Now in the critical ($\tau\to 0$) and thermodynamic limits ($N\to\infty$),
one requires for consistency that $\kappa_\tau\sim\tau^{-\gamma}$. 
Accordingly $f(N^y\tau)\sim (N^{1-y}\tau)^{-\gamma}$, and matching powers of
$N$ in equation~\ref{eq:kappa} and \ref{eq:scaleM} then yields

\begin{equation}
-2x-\gamma (1-y)=1,
\label{eq:reln}
\end{equation}
a relation that is satisfied by the assignments 
\begin{eqnarray}
x&=&\frac{\beta}{d\nu}\\
y&=&1-\frac{1}{d\nu},
\end{eqnarray}
where we have employed the hyperscaling relation $d\nu=\gamma+2\beta$ to
eliminate $\gamma$ from equation~\ref{eq:reln}.

Thus precisely at criticality, where $\tau=h=0$, equation~\ref{eq:ansatz}
can be written:

\begin{equation}
P_N({\cal M},{\cal E})\simeq a^{-1}_{\cal M}a^{-1}_{\cal E}N^{\beta/d\nu}N^{(d\nu-1)/d\nu}
\tilde{P}^\star(a^{-1}_{\cal M}N^{\beta/d\nu}({\cal M-M}_c),a^{-1}_{\cal
E}N^{(d\nu-1)/d\nu}({\cal E-E}_c)),
\label{eq:ansatz1}
\end{equation}
where $\tilde{P}^\star$ is a universal function characterising the Ising fixed
point. We note that this scaling behaviour is simply related to that for the
$\mu$VT-ensemble \cite{BRUCE2} by the substitution $L^d\to N$.
In what follows we will test the validity of equation~\ref{eq:ansatz1} via
simulations of the 3D Lennard-Jones fluid at its liquid-vapour critical point.

\section{Monte-Carlo studies}

Monte-Carlo simulations of the Lennard-Jones fluid were carried out within
the NpT-ensemble using a Metropolis algorithm \cite{WOOD}.  In
common with many other simulation studies of the model, the potential was
cutoff at a distance $r_c=2.5\sigma$, and was not shifted.  In the course of
the simulations, three separate periodic systems comprising $N=216, 363$ and $512$
particles were studied.  In order to minimise fluctuations in the acceptance
rate for volume moves, the method of Eppenga and Frenkel \cite{EPPENGA} was
employed.  Within this scheme the system performs a random walk in $\ln V$
rather than in $V$ itself, thereby automatically adjusting the step size in
the volume to be larger at lower densities than higher densities.  This
technique is particularly useful in the coexistence region region, due to
the large volume fluctuations. 

From a previous detailed simulation study of the LJ fluid in the
grand-canonical ensemble \cite{WILDING}, the critical temperature is known
to lie at $T_c\equiv 4/w_c=1.1876(3)$.  Using this value of $w$, a number of
short runs were performed in which the pressure $p$ was tuned in order to
locate approximately the coexistence curve.  Once a near-critical value of
$p$ had been obtained, a long run comprising $6\times10^6$ volume change
attempts and an equal number of displacements attempts per particle were
performed.  During this run, the histogram of the joint distribution
$P_N(\rho,u)$ was accumulated. 

\begin{figure}[h]
\vspace*{0.5 in}
\setlength{\epsfxsize}{19cm}
\centerline{\mbox{\epsffile{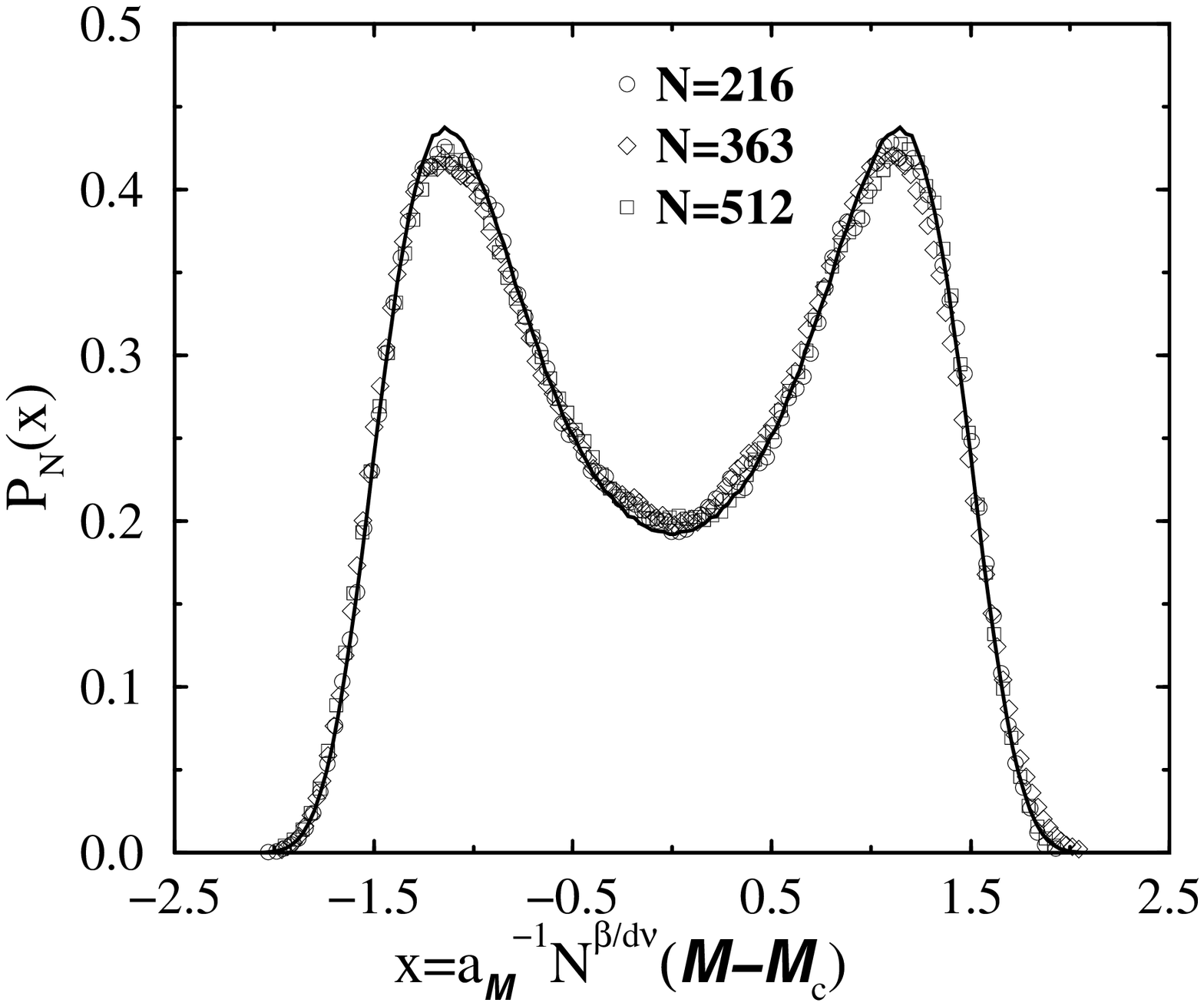}}} 

\caption{The critical scaling operator distributions $P_N({\cal M})$ for the
three system sizes $N=216, 363$ and $N=512$, expressed in terms of the
scaling variable $x=a_{\cal M}^{-1}N^{\beta/d\nu}({\cal M-M}_c)$.  The solid
line is the critical magnetisation distribution of the 3D Ising model
\protect\cite{HILFER}.}

\label{fig:poM}
\end{figure}

In order to locate more precisely the critical pressure, the ordering
operator distribution $P_N({\cal M})$ was studied for each value of $N$. 
Precisely at criticality, and for the correct choice of the field mixing
parameter $s$, $P_N({\cal M})$ is expected to be a symmetric function by
virtue of the fact that the field mixing transformation restores the Ising
symmetry of the critical fluid \cite{WILDING}.  The value of the field
mixing parameter $s$ and the pressure $p$ were therefore tuned
simultaneously using the histogram reweighting scheme \cite{FERRENBERG},
until a symmetric function was obtained.  This occurred for values
$s=-0.10(1)$ and $p=0.1093(6)$.  This value of $s$ is in accord with that
obtained previously in the GCE study of the same model \cite{WILDING}.  The
resulting form of $P_N({\cal M})$ for each system size, expressed in terms
of the scaling variable $x=a_{\cal M}^{-1}N^{\beta/d\nu}({\cal M-M}_c)$, is
plotted in figure~\ref{fig:poM}.  The non-universal scale factor $a_{\cal
M}$ was chosen so that the data for the $N=512$ system has unit variance,
while the exponent ratio $\beta/\nu$ was assigned the Ising estimate
$\beta/\nu=0.518$ of reference \cite{FERRENBERG1}.  Also shown for
comparison (solid line) is the scaled critical magnetisation distribution of
the 3D Ising model obtained in a separate study \cite{HILFER}.  Clearly the
data for each system size collapse well on to one another and are thus
consistent with the proposed scaling form.  They also agree well with the
universal form of the Ising magnetisation distribution.  We attribute the
small deviations to corrections-to-scaling associated with the rather small
numbers of particles.  These discrepancies are comparable in size and form
with those observed in reference~\cite{WILDING}.  Unfortunately, owing to
these corrections-to-scaling, as well as to the somewhat poor statistical
quality (a point to which we return in section~\ref{sec:discuss}), it was
not possible to obtain a reliable independent estimate of the exponent ratio
$\beta/\nu$. 

A procedure similar to that outlined above was also carried out for the
energy-like operator $P_N({\cal E})$.  In this case, however, there is no
symmetry requirement for the critical form of the function \cite{WILDING}. 
However, we found that by assigning the field mixing parameter $r$, the value
$r=-1.02$ found in the previous GCE study of the LJ fluid \cite{WILDING},
the function matched closely the form of the critical Ising model energy
distribution.  Again the scaling of the $P_N({\cal E})$ with $N$ was found
to be consistent with the scaling form eq.~\ref{eq:ansatz1}. 

\section{Discussions and conclusions}

\label{sec:discuss}

The aim of this work was to develop a FSS formalism for simulations of
near-critical fluids at constant pressure.  A scaling form was proposed and
tested for the Lennard-Jones fluid, whose critical point parameters are
known to high precision.  Good consistency with the scaling prediction was
found.  Additionally, it was shown that the forms of the critical scaling
operator distributions are the {\em same} as those for the $\mu$VT-ensemble
(and hence the canonical ensemble of the critical Ising model), despite the
very different nature of the simulation ensembles.  As is well known,
finite-size effects {\em may} differ in various ensembles of statistical
mechanics: the microcanonical ensemble, canonical and grand canonical
ensembles of fluids are known to have distinct finite-size properties, and
are characterized by different scaling functions.  Our interpretation of the
equivalence of the constant-NpT and constant-$\mu VT$ ensemble scaling
functions, is that in the thermodynamic limit, $p=p(T,\mu)$, and therefore
it is immaterial whether one uses $\mu T$ or $p T$ as the pair of given
intrinsic thermodynamic variables.  All that matters is that both ensembles
have a single extensive variable ($N$ or $V$).  Different scaling properties
are obtained when two extensive variables are used, such as in the
constant-NVT (canonical) ensemble of fluids.  It follows that the GCE-based
FSS techniques developed in reference \cite{WILDING} for accurately locating
fluid critical point parameters, carry over directly to the NpT-ensemble. 
It is hoped that use of this latter ensemble will prove beneficial in
situations where the GCE is highly inefficient, such as for long-chain
polymer or electrolyte models.  We add the further comment that finite-size
scaling with $N$ rather than $V$ is well known in mean-field spin systems,
when every spin interacts with every other spin, and geometry and space have
no meaning \cite{BOLET}. 

While use of the NpT-ensemble is likely to be much more efficient
than the GCE in cases where particle insertions have a low acceptance
probability, we believe that for simple fluids, use of the grand canonical
ensemble is much to be preferred.  The present study of the LJ fluid
consumed circa $300$ hours CPU time on a Cray-YMP, using a vectorized
program.  By contrast, the previous GCE study of the same model was
performed with considerably less computational effort using workstations. 
Moreover, it was possible to study much larger systems (comprising up to
$\sim 1800$ particles), with considerably greater statistical accuracy than
attainable here.  The reasons for this difference in efficiency seem to be
manifold.  Owing to the fluctuating volume, it is not possible to easily
implement a cell structure for efficiently locating neighbouring particles
within the cutoff range.  A calculation of all particle separations prior to
a trial volume transition involves an $O(N^2)$ operation, which can only be
avoided if no potential cutoff is employed \cite{ALLEN1}.  By contrast, use
of a cell structure in the GCE leads to a $O(N)$ growth in computational
complexity.  A second disadvantage of the NpT-ensemble, is that
both volume changes and particle moves must be performed.  For each particle
move, two partial energy calculations are required, while for a volume
change (with finite interaction cutoff), a total energy calculation is
required.  By contrast, only particle insertions need be implemented in a GCE
scheme \cite{WILDING}, each of which requires only a single partial energy
calculation.  Additionally it seems likely that for a given
$\langle\rho\rangle$ the random walk in $V$ required to pass from one phase
to the other at coexistence is longer than that in $N$ for the GCE, even if
one employs a dynamically adjusting volume step size such as that of
reference \cite{EPPENGA}.  Consequently the correlation time in the
NpT-ensemble is considerably greater than that of the GCE. 

\begin{figure}[h]
\vspace*{0.5 in}
\setlength{\epsfxsize}{19cm}
\centerline{\mbox{\epsffile{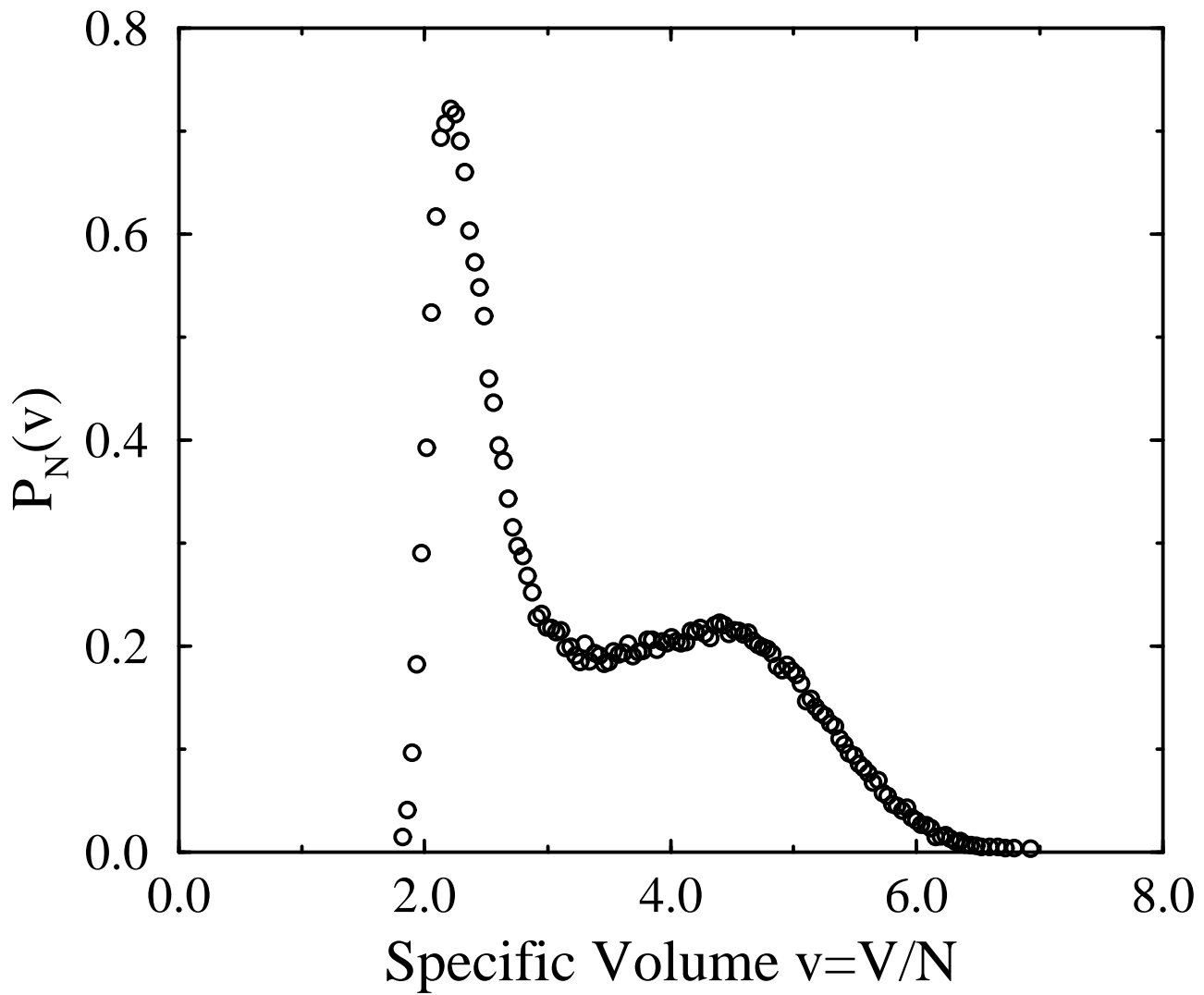}}} 

\caption{The specific volume distribution $P_N(v)$ for the $N=512$ system,
measured at the assigned values of the critical parameters.}

\label{fig:pV}
\end{figure}

Finally we note that another recent study has also attempted to apply
finite-size scaling methods to the LJ fluid in the NpT-ensemble
\cite{HUNTER}.  In this study the authors focused on the distribution of
the specific volume $v=V/N$ and attempted to locate the critical point
by requiring scale invariance in $P_N(v)$, subject to the requirement
that $P_N(v)$ has two peaks of {\em equal} height at coexistence. 
However, as we have shown in this work, the volume distribution is not
the scaling counterpart for fluids of the Ising magnetisation, rather it
is the operator ${\cal M}$.  Moreover, the limiting (large $N$) critical
point form of $P_N(v)$ does not have two peaks of equal heights, and is
in fact highly asymmetic in form, as figure~\ref{fig:pV} shows.  We
therefore ascribe the rather low accuracy in the estimates for $T_c$ and
the difficulties in obtaining the surface tension exponent in reference
\cite{HUNTER} to this incorrect choice of the scaling operator. 

\subsection*{Acknowledgements}

Helpful correspondence with A.D.  Bruce and G.  Orkoulas is acknowledged. 
NBW is supported by the Max-Planck Institute f\"{u}r Polymerforschung,
Mainz.

\end{document}